\documentclass[twocolumn,10pt,prb,aps,amssymb,amsmath,superscriptaddress]{revtex4}

\usepackage{color,graphicx,ulem}
\usepackage{bm}

\newcommand{\vth}[1][s]{\ensuremath{v_{\mathrm{th}_#1}}}
\newcommand{\pd}[2]{\ensuremath{ \frac{\partial #1} {\partial #2} } }
\newcommand{\gyror}[1]{\ensuremath{ {\left< #1 \right>}_\mathbf{r}}}
\newcommand{\gyroR}[2][s]{\ensuremath{{\left< #2 \right>}_\mathbf{R_\mathrm{#1}}}}

\newcommand{\unit}[1]{\bm{\hat{#1}}}

\renewcommand{\eqref}[1]{Eq.\ (\ref{#1})}
\newcommand{\eqsref}[2]{Eqs.\ (\ref{#1}) and (\ref{#2})}
\newcommand{\eqsdash}[2]{Eqs.\ (\ref{#1})--(\ref{#2})}
\newcommand{\Secref}[1]{Section \ref{#1}}
\newcommand{\secref}[1]{Sec.\ \ref{#1}}
\newcommand{\apref}[1]{Appendix \ref{#1}}

\renewcommand{\vth}{\ensuremath{ v_{\mathrm{th}} } }
\renewcommand{\gyroR}[1]{\ensuremath{{\left< #1 \right>}_{\bm{R}}}}
\newcommand{\Cgk}{C_{\text{GK}}}
\newcommand{\hk}{h_{\bm{k}}}

\begin{document}

\eprint{arXiv:0808.1300}
\title{Linearized model Fokker--Planck collision operators for gyrokinetic simulations.\\ I. Theory}

\author{I.\ G.\ Abel}
\email{ian@utumno.org.uk}
\affiliation{Plasma Physics Group, Blackett Laboratory, Imperial College, London SW7 2AZ, UK
}
\affiliation{Euratom/UKAEA Fusion Association, Culham Science Centre, Abingdon OX14 3DB, UK
}
\author{M.\ Barnes}
\email{mabarnes@umd.edu}
\affiliation{Department of Physics, IREAP and CSCAMM, University of Maryland, College Park, Maryland 20742-3511
}
\author{S.\ C.\ Cowley}
\affiliation{Plasma Physics Group, Blackett Laboratory, Imperial College, London SW7 2AZ, UK
}
\affiliation{Euratom/UKAEA Fusion Association, Culham Science Centre, Abingdon OX14 3DB, UK
}
\author{W.\ Dorland}
\affiliation{Department of Physics, IREAP and CSCAMM, University of Maryland, College Park, Maryland 20742-3511
}
\author{A.\ A.\ Schekochihin}
\affiliation{Plasma Physics Group, Blackett Laboratory, Imperial College, London SW7 2AZ, UK
}

\date{\today}

\begin{abstract}
A new analytically and numerically manageable model collision operator is developed specifically for turbulence simulations. The like-particle collision operator includes both pitch-angle scattering and energy diffusion and satisfies the physical constraints required for collision operators: it conserves particles, momentum and energy, obeys Boltzmann's $H$-theorem (collisions cannot decrease entropy), vanishes on a Maxwellian, and efficiently dissipates small-scale structure in the velocity space. The process of transforming this collision operator into the gyroaveraged form for use in gyrokinetic simulations is detailed. The gyroaveraged model operator is shown to have more suitable behavior at small scales in phase space than previously suggested models. Model operators for electron-ion and ion-electron collisions are also presented. 
\end{abstract}
\maketitle

\section{Introduction}
\label{sec_intro}

It has long been known that in many turbulent systems the differences between vanishingly small dissipation and neglecting dissipation completely are striking, and that this can be linked theoretically to the non-interchangeability of limits $t \rightarrow \infty$ and $\nu \rightarrow 0$, where $\nu$ is, e.g., viscosity, resistivity or collision frequency. Physically, the dissipation is important in turbulence for the following reason. The fundamental property of turbulence is to transfer energy from scales at which it is injected into the system to scales where it is dissipated, leading to heating. When the dissipation coefficients are small, the system has to generate very fine-scale fluctuations in order to transfer the energy to scales at which dissipation becomes efficient. 

Because of Boltzmann's $H$-theorem,\cite{Boltzmann} dissipation (meaning any effect that leads to irreversible heating) in kinetic plasmas is ultimately collisional, so the transfer of energy generally occurs in phase space --- i.e., both in position and velocity space (see extended discussion of energy cascade in plasma turbulence in Ref.\ \onlinecite{schekcrete} and references therein). There are a number of specific mechanisms, both linear and nonlinear, that give rise to phase-space mixing.\cite{hammett:2052,dorland:812,Krommes2,Krommes1,watanabe:1476,Tome,schekcrete} It is the resulting large gradients in the velocity space that eventually bring collisions into play however small the collision frequency is (such small-scale velocity-space structure has, e.g., been found and explicitly measured in gyrokinetic simulations\cite{watanabe:1476,howesdpp,numerics,tatsuno}). Thus, in any plasma turbulence simulation, some effective collisionality has to be present in order to smooth the small-scale structure in velocity space. 

Besides velocity-space smoothing, there is another key reason why collisions must be included. Collisions, through the dissipation of small-scale fluctuations in phase space, provide the physical link between irreversible plasma heating (macroscopic transport) and turbulence which enables the system to converge to a statistically steady state. Although it is possible for a collisionless simulation to temporarily achieve a quasi-steady state in macroscopic quantities, achieving a true steady state in the long-time limit requires some form of dissipation.\cite{Krommes1} While many simulations in plasma physics and neutral fluid dynamics have used numerical dissipation, such as simple hyperdiffusion (or more sophisticated subgrid turbulence models in Large Eddy Simulations), to provide the dissipation needed for steady state, it is important to also be able to carry out direct numerical simulations, where the physical dissipation processes are explicitly resolved.  This provides a valuable cross-check on simulations with numerical dissipation, and is useful as a standard by which one could search for optimal subgrid models.

Let us explain in more detail why collisions are important for achieving the steady state. Consider the ``$\delta f$ kinetics,'' i.e., assume that it is physically reasonable to split the distribution function into a slowly (both spatially and temporally) varying equilibrium part and a rapidly varying fluctuating part: $f= F_0+\delta f$. We further assume that $F_0$ is a Maxwellian distribution, $F_0=(n_0/\pi^{3/2}\vth^3)\exp(-v^2/\vth^2)$, where $n_0$ is density, $\vth=(2T_0/m)^{1/2}$ the thermal speed, $T_0$ temperature and $m$ the particle mass. This will be the case if collisions are not extremely weak (for the weakly collisional formulation of $\delta f$ gyrokinetics, see Ref.\ \onlinecite{howes2006agb}). One can show that the fundamental energy balance governing the evolution of the turbulent fluctuations is\cite{Krommes2,sugama1996tpa,watanabe:1476,hallatschek,howes2006agb,Tome,scott,schekcrete}
\begin{equation}
\label{balance}
\begin{split}
&\frac{d}{dt} \left( -\sum_s T_{0s}\delta S_s + U \right) =\\
&\qquad P + \sum_s\iint\frac{T_{0s}\delta f_s}{F_{0s}}\,C[\delta f_s]
d \bm{v} d \bm{r},
\end{split} 
\end{equation}
where $s$ is the species index, 
$\delta S = -\iint d\bm{r}d\bm{v}\,\delta f^2/2F_0$ is the entropy of the fluctuations, 
$U=\int d\bm{r}\,(E^2+B^2)/8\pi$ is the energy of the (fluctuating) electromagnetic field, 
$P$ is the input power (energy source of the turbulence), 
and $C[\delta f]$ is the linearized collision operator. 
In many types of plasma turbulence studied in fusion contexts, the input power 
$P$ is proportional to the heat flux and it is the parameter dependence of 
the mean value of this quantity in the statistically stationary state 
that is sought as the principal outcome of the simulations. 
We can see immediately from the above equation that collisions (or some form of dissipation) 
are required to achieve such a steady state (as has been shown in numerical 
simulations\cite{Krommes1,watanabe:3659,watanabe:1476,CandyWaltz}) 
and that in this steady state, $P$ must be balanced on the average by the dissipation 
term.

A key property of the collision operator required for this transfer of 
energy from turbulence to the equilibrium distribution to work correctly and, therefore, for 
the heat fluxes to converge to correct steady-state values, is that the collision term in 
\eqref{balance} must be negative-definite: 
\begin{equation}
\label{Htheorem}
\iint \frac{\delta f}{F_{0}}\, C[\delta f] d\bm{v}d\bm{r} \le 0. 
\end{equation}
This ensures that heating is irreversible and that collisions cannot decrease entropy, the latter being the statement of Boltzmann's $H$-theorem.\cite{Boltzmann} 
While the heat fluxes might not be sensitive to the exact form of a model collision operator (within some range of models) at low collision frequency, any spurious sink of entropy may adversely affect the balance between turbulent fluxes and dissipation. Therefore it is clearly preferable that a model collision operator respect the $H$-theorem, which has important physical consequences. Preserving the $H$-theorem may be even more important at higher collision frequencies. 

In view of the above discussion, we can formulate a reasonably restrictive set of criteria for any model collision operator: providing dissipation at small scales, obeying the $H$-theorem [\eqref{Htheorem}], and also, obviously, conserving particle number, momentum, energy, and vanishing on a (local, perturbed) Maxwellian distribution.  Whilst these properties are analytically convenient, for numerical simulations the operator should also be efficiently implementable and carry these properties (at least approximately) over to the numerical scheme.

	The effect of small angle Coulomb collisions on an arbirtrary distribution function was originally calculated by Landau.\cite{landau1936kgf} In the $\delta f$ kinetics we would naturally consider the linearized Landau operator.\cite{helander2002ctm} However, it is sufficiently complex that a direct numerical convolution evaluation of it would exceed the limits on numerical resources that can be realistically expended on modelling the collisional physics of predominantly collisionless plasmas. 
Consequently several simplified model collision operators have been developed, both for analytical and computational convenience, that try to capture the qualitative essence, if not the quantitative detail, of the physics involved.\cite{rutherford1970ele,hirshman1976afp,catto1977col} This course of action is, indeed, eminently sensible: from \eqref{balance}, it seems plausible that, at least as far as calculating integral characteristics such as the turbulent fluxes is concerned, neither the exact functional form of the collision operator (provided it satisfies the criteria discussed above) nor the exact value of the collision frequency (provided it is sufficiently small) should be important. All we need is a physically reasonable dissipation mechanism.

For these purposes, it has often been deemed sufficient to use the pitch-angle-scattering (Lorentz) operator, sometimes adjusted for momentum conservation.\cite{rutherford1970ele,helander2002ctm} However, in kinetic turbulence, there is no reason that small-scale velocity-space structure should be restricted to pitch angles. In fact, standard phase-mixing mechanisms applied to gyrokinetics produce structure in $v_\parallel$,\cite{hammett:2052,watanabe:1476} and there is also a nonlinear gyrokinetic phase mixing that gives rise to structure in $v_\perp$, which may be an even faster and more efficient process.\cite{dorland:812,Tome,schekcrete} Thus, {\it a priori} one expects to see small scales both in the pitch angle and in the energy variables ($\xi$ and $v$). It has, indeed, been confirmed in simulations\cite{numerics} that with only Lorentz scattering structure rapidly forms at the grid scale in energy. Thus a numerically suitable model collision operator should include energy diffusion.\footnote{Energy diffusion is also known to be important in higher collisionality regimes, leading to significant effects on neoclassical transport and on various instability mechanisms (see Ref.~\onlinecite{numerics} and references therein).} 

In this paper, we propose such an operator (other operators including energy diffusion have been previously suggested;\cite{hirshman1976afp,catto1977col} we include a detailed comparison of our operator with these in \apref{ap_prev}). Our model operator for like-particle collisions, including both pitch-angle scattering and energy diffusion and satisfying all of the physical constraints discussed above, is given in \secref{sec_newop} (the proof of the $H$-theorem for it is presented in \apref{ap_Hth}). In \secref{sec_gk}, it is converted (gyroaveraged) into a form suitable for use in gyrokinetic simulations --- a procedure that produces some nontrivial modifications. In \secref{sec_ei}, we explain how interspecies collisions can be modelled in gyrokinetic simulations to ensure that such effects as resistivity are correctly captured. \Secref{sec_summary} contains a short summary and a discussion of the consequences of the work presented here. 

The anlytical developments presented in this paper form the basis for the numerical implementation of collisions in the publicly available gyrokinetic code \verb#GS2#. This numerical implementation, as well as a suite of numerical tests are presented in the companion paper, Ref.\ \onlinecite{numerics} (henceforth Paper II). 

\section{A New Model Collision Operator}
\label{sec_newop}

In this section, we present a new model collision operator for like-particle collisions that 
satisfies the criteria stated above. The interpecies collisions will be considered in \secref{sec_ei}.

Let us start by introducing some standard notation. In discussing collision operators on phase space, we shall denote by $\bm{r}$ the position variable in physical space and use $(v,\xi,\vartheta)$ coordinates in velocity space, where $v = |\bm{v}|$ is the energy variable, $\xi = v_\parallel/v$ is the pitch-angle variable, and $\vartheta$ the gyroangle about the equilibrium magnetic field. One can easily adapt the operators presented here to unmagnetized plasmas, but as we are interested in gyrokinetic plasmas, we shall concentrate on the strongly magnetized case. Taking the notation of Ref.~\onlinecite{helander2002ctm} as the standard, we introduce the normalized velocity variable $x=v/\vth$ and a set of velocity-dependent collision frequencies for like-particle collisions: 
\begin{eqnarray}
\label{nuD_def}
\nu_{D}(v) &=& \nu\, \frac{\Phi( x ) - G( x ) }{ x^3}, \\
\label{nus_def}
\nu_s(v) &=& \nu\,\frac{4G( x ) }{  x},\\
\label{nupar_def}
\nu_{\parallel}(v) &=& \nu\, \frac{2G( x ) }{ x^3} = \frac{1}{2x^2}\,\nu_s,\\
\label{dnu_def}
\Delta\nu(v) &=& \nu_D - \nu_s = \frac{1}{2 v^3 F_0}\pd{}{v}\,v^4\nu_\parallel F_0,\\
\label{nuE_def}
\nu_{E}(v) &=& -2 \Delta\nu - \nu_\parallel
= - \frac{1}{v^4 F_0}\pd{}{v}\,v^5\nu_\parallel F_0,
\end{eqnarray}
where $\Phi(x) = (2/\sqrt{\pi})\int^x_0 e^{-y^2} dy$ is the error function, 
$G(x) = [\Phi(x)-x\Phi'(x)]/2x^2$ is the Chandrasekhar function, and
$\nu = \sqrt{2}\pi n_0 q^4\ln\Lambda\, T_0^{-3/2}m^{-1/2}$ is the dimensional like-particle 
collision frequency (here $\ln\Lambda$ is the Coulomb logarithm and $q$ is the particle charge). 
Note that the two differential identities given in \eqsref{dnu_def}{nuE_def} will 
prove very useful in what follows. 

If one wishes to construct a model linearized collision operator, the following general form constitutes a natural starting point
\begin{equation}
C[\delta f] = \pd{}{\bm{v}} \cdot \left[\hat D(\bm{v})\cdot \pd{}{\bm{v}} 
\frac{\delta f}{F_0}\right] + P[\delta f](\bm{v}) F_{0},
\end{equation}
where the first term is the ``test-particle'' collision operator and the second term the 
``field-particle'' operator. Most model operators can be obtained by picking a suitably simple form for the velocity-space diffusion tensor $\hat D$ and the functional $P$, 
subject to the constraints that one chooses to impose on the model operator. 

In constructing our model 
operator, we retain the exact form of $\hat D$ for the linearized Landau collision 
operator:\cite{helander2002ctm} 
\begin{equation}
C[\delta f] = \nu_D L[\delta f] + \frac{1}{v^2} \pd{}{v}\left(\frac{1}{2} v^4 \nu_\parallel F_{0} \pd{}{v}\frac{\delta f}{F_0} \right) + P[\delta f](\bm{v}) F_{0},
\end{equation}
where we have explicitly separated the energy-diffusion part (the second term) and 
the angular part (the first term), which includes pitch-angle scattering and 
is described by the Lorentz operator: 
\begin{equation}
\label{def_L}
L[\delta f] = \frac{1}{2}\left[ \pd{}{\xi}(1-\xi^2)\pd{\delta f}{\xi} + 
\frac{1}{ 1-\xi^2}\pd{^2\delta f}{\vartheta^2}\right].
\end{equation}

Our modelling choice is to pick $P$ to be of the form
\begin{equation}
\label{pres}
P[\delta f](\bm{v}) = \nu_s\, \frac{2\bm{v} \cdot \bm{U}[\delta f]}{\vth^2}
+ \nu_E\, \frac{v^2}{\vth^2}\, Q[\delta f].
\end{equation}
One can view this prescription as first expanding $P$ in spherical harmonics (one can easily show that they are eigenfunctions of the full field-particle operator), reataining only the first two terms, and then arbitrarily factorizing the explicit $v$ and $\delta f$ dependence of each harmonic. The functionals $\bm{U}[\delta f]$ and $Q[\delta f]$ are mandated to have no explicit velocity dependence. In this ansatz the $v$ dependence is chosen so that the final operator is self adjoint and also to ensure automatic particle conservation by the field-particle operator: $\int P[\delta f](\bm{v}) F_0\,d\bm{v}=0$. Indeed the first term in \eqref{pres} gives a vanishing contribution to this integral because it is proportional to $\bm{v}$, and so does the second term because of the differential identity given in \eqref{nuE_def}.
The functionals $\bm{U}[\delta f]$ and $Q[\delta f]$ are now uniquely chosen so as to ensure that the model operator conserves momentum and energy: a straightforward calculation gives
\begin{align}
\label{def_U}
\bm{U}[\delta f] &= \frac{3}{2}\int \nu_s \bm{v} \delta f\, d\bm{v} \left/ 
\int \left(v/\vth\right)^2 \nu_s F_{0}\, d\bm{v}, \right.\\
\label{def_Q}
Q[\delta f] &= \int v^2 \nu_E \delta f\, d\bm{v} \left/ 
\int v^2 \left(v/\vth\right)^2 \nu_E F_{0}\, d\bm{v}.\quad\right. 
\end{align} 
These are in fact just the standard correction expressions used for the model pitch-angle-scattering operator\cite{rutherford1970ele,helander2002ctm} and for more complex operators including energy diffusion.\cite{hirshman1976afp} 

Note that the numerical implementation of our collision operator documented in Paper II\cite{numerics} achieves exact satisfaction of the conservation laws by choosing the discretization scheme that exactly captures the differential identities \eqsref{dnu_def}{nuE_def} and the double integration by parts needed in deriving \eqsref{def_U}{def_Q}.

To summarize, we now have the following model operator for like particle collisions:
\begin{widetext}
\begin{equation}
\label{model}
C[\delta f] = \frac{\nu_D}{2}\left[ \pd{}{\xi}(1-\xi^2)\pd{\delta f}{\xi} + 
\frac{1}{ 1-\xi^2}\pd{^2\delta f}{\vartheta^2}\right]
+ \frac{1}{v^2}\pd{}{v} \left( \frac{1}{2} v^4\nu_\parallel F_{0} \pd{}{v}\frac{\delta f}{F_0}\right) 
+ \nu_s\,\frac{2\bm{v} \cdot \bm{U}[\delta f] }{\vth^2} F_{0} 
+ \nu_E\,\frac{v^2}{\vth^2}\, Q[\delta f] F_{0},
\end{equation}
\end{widetext}
where the functionals $\bm{U}[\delta f]$ and $Q[\delta f]$ are given by \eqsref{def_U}{def_Q}. The modelling choice of the field-particle operator that we have made [\eqref{pres}] means that, in order to compute our collision operator, we have only to calculate definite integrals over the entirety of the velocity space --- a significant simplification in terms of computational complexity and ease of use in numerical simulations (compared to computing convolutions over velocity space, see Paper II\cite{numerics}). 

As we have shown above, our operator conserves particles, momentum and energy by construction.  It is also not hard to see that it vanishes precisely when $\delta f / F_0 = 1, \bm{v}, v^2$ and linear combinations thereof, i.e. if $\delta f$ is a perturbed Maxwellian.  From this and the fact that the operator is self adjoint, it can be easily shown that the operator {\it only} conserves particles, momentum and energy and that no spurious conservation laws have been introduced by our model.  Because the model we have chosen retains the exact Landau test-particle operator , it provides velocity-space diffusion both in energy and in pitch angle and will thus efficiently dissipate small-scale structure in velocity space.  Finally, our model satisfies the $H$-theorem --- this is proved in \apref{ap_Hth}. 

This operator thus fulfills the criteria set forth in \secref{sec_intro} to be satisfied by any physically reasonable model operator. We now proceed to convert this operator into a form suitable for use in gyrokinetics.

\section{Collisions in Gyrokinetics}
\label{sec_gk}

The gyrokinetic theory is traditionally derived for a collisionless plasma.\cite{frieman1982nge,brizard2007} However, as we have argued in \secref{sec_intro}, even when the collision frequency is small, collisions must be included in order to regularize the phase space and to ensure convergence of fluxes to statistically stationary values. Mathematically, collisions can be included in gyrokinetics if the collision frequency is formally ordered to be comparable to the fluctuation frequency,\cite{howes2006agb}  $\nu\sim\omega\sim k_\parallel\vth$ --- the weakly collisional limit (collisionality larger than this leads simply to fluid equations).  In practice, the collision frequency tends to be smaller than typical fluctuation frequencies, but this need not upset the formal ordering as long as it is not too small: the cases $\nu\gg\omega$ and $\nu\ll\omega$ can be treated as subsidiary limits.\cite{Tome}

Under the formal ordering $\nu\sim\omega$
, it is possible to show that the equilibrium distribution function (lowest order in the gyrokinetic expansion) is a Maxwellian\cite{howes2006agb} and the full distribution function can be represented as 
\begin{equation}
f = \left(1 - \frac{q \varphi}{T_{0}}\right) F_{0} + h(t,\bm{R},\mu,\varepsilon), 
\end{equation}
where $F_{0}$ is a Maxwellian, $\varphi$ the electrostatic potential (a fluctuating quantity) and $h$ the (perturbed) distribution function of the paticle guiding centers.  Here $\varepsilon=mv^2/2$ is the particle energy, $\mu=mv_\perp^2/2B_0$ the first adiabatic invariant, $B_0$ the strength of the equilibrium magnetic field,  $\bm{R} = \bm{r} - \bm{\rho} = \bm{r} - \unit{b}\times\bm{v} / \Omega$ the guiding center position, $\Omega$ the cyclotron frequency, and $\unit{b}=\bm{B}_0/B_0$. The gyrokinetic equation, written in general geometry and including the collision operator is then
\begin{eqnarray}
\nonumber
&&\pd{h}{t} + (v_\parallel\unit{b} + \bm{v}_D)\cdot\pd{h}{\bm{R}} 
+ \frac{c}{B_0}\left\{\gyroR{\chi},h\right\}\\ 
&&\quad= -q\,\pd{F_0}{\varepsilon}\pd{\gyroR{\chi}}{t} 
+ \frac{c}{B_0}\left\{F_0,\gyroR{\chi}\right\} + \Cgk[h],\quad
\label{GK_eqn}
\end{eqnarray}
where $\chi = \varphi - \bm{v}\cdot\bm{A}/c$ the gyrokinetic potential, 
$\gyroR{\chi} = (1/2\pi) \int \chi(\bm{R}+\bm{\rho})\,{d}\vartheta$ is 
an average over gyroangles holding $\bm{R}$ fixed (the ``gyroaverage''), 
$\bm{v}_{D}$ is the guiding center drift velocity. 

The gyrokinetic collision operator $\Cgk[h]$ is the gyroaverage of the linearized collision operator. 
The latter acts on the perturbed distribution $h$ holding the particle position $\bm{r}$
(not the guiding center $\bm{R}$!) fixed. This nuance must be kept in mind whilst working 
out the explicit form of $\Cgk[h(\bm{R})]$ from the unaveraged linearized operator 
$C[h(\bm{r}-\bm{\rho})]$. 

Let us restrict our consideration to local simulations, which are carried out in a flux tube of long parallel extent, but short perpendicular extent. In such simulations, one assumes that the equilibrium profiles are constant across the tube, but have non-zero gradients across the tube so as to keep all the appropriate drifts and instabilities. This permits one to use periodic boundary conditions and perform the simulations spectrally perpendicular to field lines.\cite{beer:2687} Thus
\begin{equation}
h = \sum\limits_{\bm{k}} e^{i\bm{k}\cdot\bm{R}} h_{\bm{k}}(l,v,\mu),
\end{equation}
where $l$ is a coordinate along the field line and the Fourier transform is understood to be only with respect to the perpendicular components of $\bm{R}$, i.e., $\bm{k}\equiv\bm{k}_\perp$. Treating the perpendicular coordinates spectrally confines all dependence on the gyroangle $\vartheta$ to the exponent, thus we can compute the gyroangle dependence explicitly and carry out 
the gyroaveraging of the collision operator in a particularly 
transparent analytical way:\cite{catto1977col,Tome} 
\begin{widetext}
\begin{equation}
\Cgk[h] = \gyroR{C\left[\sum_{\bm{k}} e^{i\bm{k}\cdot\bm{R}}\hk\right]}\\ 
= \sum_{\bm{k}} \gyroR{e^{i\bm{k}\cdot\bm{r}}C[e^{-i\bm{k}\cdot\bm{\rho}}\hk]}\\
= \sum_{\bm{k}} e^{i\bm{k}\cdot\bm{R}}\gyroR{e^{i\bm{k}\cdot\bm{\rho}}
C[e^{-i\bm{k}\cdot\bm{\rho}}\hk]},
\label{Cgk_avg}
\end{equation}
\end{widetext}
where $\bm{\rho}=\unit{b}\times\bm{v}_\perp/\Omega$. 
Thus, in Fourier space
\begin{equation} 
\label{Cgk_gen}
\Cgk[\hk]= \left<e^{i\bm{k}\cdot\bm{\rho}}C[e^{-i\bm{k}\cdot\bm{\rho}}\hk]\right>,
\end{equation} 
where $\left<\dots\right>$ refers to the explicit averaging over the $\vartheta$ dependence. 
Some general properties of this operator are discussed in Appendix B of Ref.~\onlinecite{Tome}. 

We now apply the general gyroaveraging formula \eqref{Cgk_gen} 
to our model operator given by \eqref{model}. 
The gyrokinetic transformation of variables 
$(\bm{r},v,\xi,\vartheta)\to(\bm{R},\mu,\varepsilon,\vartheta)$ 
mixes position and velocity space. 
However, in the collision operator, to the lowest order 
in the gyrokinetic expansion, we can neglect the spatial dependence of $\mu$ that comes via 
the equilibrium magnetic field $B_0(\bm{r})$ and thus use the $(v,\xi)$ velocity variables.  
After some straightforward algebra, which involves converting velocity derivatives at constant $\bm{r}$ 
to those at constant $\bm{R}$ and evaluating the arising gyroaverages as detailed in 
\apref{ap_gyroavg}, we arrive at the following model gyrokinetic collision operator
\begin{widetext}
\begin{eqnarray}
\nonumber
\Cgk[\hk] &=& \frac{\nu_D}{2}\,\pd{}{\xi}(1-\xi^2)\pd{\hk}{\xi}
+ \frac{1}{v^2}\pd{}{v} 
\left( \frac{1}{2} v^4\nu_\parallel F_{0} \pd{}{v}\frac{\hk}{F_0}\right)
- \frac{1}{4}\left[\nu_D(1+\xi^2) + \nu_\parallel(1-\xi^2)\right]\frac{v^2}{\vth^2}\,
k_\perp^2\rho^2 \hk\\
&& + 2\nu_s\,\frac{v_\perp J_1(a)U_\perp[\hk] + v_\parallel J_0(a)U_\parallel[\hk]}{\vth^2}\,F_0 
+ \nu_E\,\frac{v^2}{\vth^2}J_0(a)Q[\hk]F_0,
\label{gyroav}
\end{eqnarray}
where $\rho=\vth/\Omega$ is the thermal Larmor radius (not to be confused with the 
velocity-dependent $\bm{\rho}$), $a = k_\perp v_\perp / \Omega$, 
$J_0$ and $J_1$ are Bessel functions and 
\begin{eqnarray}
\label{Uperp}
U_\perp[\hk] &=& \frac{3}{2}\int \nu_s v_\perp J_1(a) \hk\, d\bm{v} \left/ 
\int { \left(v/\vth\right)^2 \nu_s F_{0}\, d\bm{v} }, \right. \\
\label{Upar}
U_\parallel[\hk] &=& \frac{3}{2}\int \nu_s v_\parallel J_0(a) \hk\, d\bm{v} \left/ 
\int { \left(v/\vth\right)^2 \nu_s F_{0}\, d\bm{v} }, \right. \\
\label{Qhk}
Q[\hk] &=& \int v^2 \nu_E J_0(a) \hk\, d\bm{v} \left/ 
\int { v^2 \left(v/\vth\right)^2 \nu_E F_{0}\, d\bm{v} }. \right.
\end{eqnarray}
\end{widetext}
Note that since the position and velocity space are mixed by the gyrokinetic 
transformation of variables, $\bm{R}=\bm{r}-\bm{\rho}$, the collision operator now 
contains not just pitch-angle and $v$ derivatives but also a spatial perpendicular 
``gyrodiffusion'' term. 

It important to make sure that the operator we have derived behaves in a physically 
sensible way in the long- and short-wavelength limits. When $k_\perp\rho\ll1$, 
all the finite-Larmor-radius (FLR) effects, including the gyrodiffusion, disappear and we 
end up with pitch-angle scattering and energy diffusion corrected for energy and 
parallel momentum conservation --- the drift-kinetic limit. 
In the opposite limit, $k_\perp\rho\gg1$, we can estimate the behavior of our operator 
by adopting the scaling of the velocity derivatives based on the nonlinear perpendicular 
phase mixing mechanism for gyrokinetic turbulence proposed in Ref.~\onlinecite{schekcrete}: 
this produces velocity-space structure with characteristic gradients 
$\vth\partial/\partial v_\perp \sim k_\perp\rho$ (see also Refs.~\onlinecite{dorland:812,Tome}). 
With this estimate, we see that all the field-particle terms in the operator are subdominant 
by a factor of $(k_\perp \rho)^{-3}$. 
Thus the operator reduces to the gyrokinetic form of the test-particle Landau operator 
in this limit. All diffusive terms are also equally large in this scaling, 
supporting our supposition that energy diffusion needs to be included.
These considerations give us some confidence that we correctly model the diffusive aspects of the collisional physics in a short-wavelength turbulent regime. Indeed, if one applies the same estimates to the full linearized Landau operator, the Rosenbluth potentials of the perturbation are small when $k_\perp \rho\gg1$ because they are integrals of a rapidly oscillating function, so the dominant effect does, indeed, come entirely from the test-particle part of the operator. 

The gyrokinetic $H$-theorem, which has to be satisfied in order for heating and transport to be correctly calculated, is given by\cite{howes2006agb,Tome} 
\begin{equation}
\iint \frac{h}{F_0}\,\Cgk[h]\,d\bm{R}d\bm{v}\le 0.
\end{equation}
The gyrokinetic collision operator given by \eqref{gyroav} respects this inequality, 
as can either be shown directly from \eqref{gyroav} (analogously to 
the proof in \apref{ap_Hth})
or inferred from \eqref{Htheorem} by transforming to gyrokinetic variables.
The operator also manifestly diffuses small-scale 
structure both in velocity and in (perpendicular) position space. 

How to express the conservation-law tests upon this operator is a somewhat subtler question.
This is because after the gyroaveraging has been done, 
one cannot explicitly separate the position- and velocity-space dynamics 
in the gyrokinetic phase space. However, it is still possible to express the evolution 
of particles, momentum and energy as local conservation laws. 
Let us take the velocity moments of the gyrokinetic equation \eqref{GK_eqn}
corresponding to these conserved quantities. These are evaluated at constant position $\bm{r}=\bm{R}+\bm{\rho}$.
Defining $\gyror{\dots}$ as the average over gyroangles while holding $\bm{r}$ fixed 
and using \eqref{Cgk_avg}, we arrive at the following evolution 
equation for the conserved moments: 
\begin{align}
\nonumber
&\pd{}{t}\int \begin{pmatrix} 1\\ \bm{v}\\ v^2\end{pmatrix} 
\delta f(t,\bm{r},\bm{v}) d\bm{v} 
+ \bm{\nabla}\cdot\bm{\hat\Gamma}^{(0)}\\ 
\nonumber
&= \int\gyror{\begin{pmatrix} 1\\ \bm{v}\\ v^2\end{pmatrix}\Cgk[h]} d\bm{v}\\ 
\nonumber
&= \sum_{\bm{k}} e^{i\bm{k}\cdot\bm{r}}\int
\left<\begin{pmatrix} 1\\ \bm{v}\\ v^2\end{pmatrix}e^{-i\bm{k}\cdot\bm{\rho}} \right>
\left<e^{i\bm{k}\cdot\bm{\rho}}C[e^{-i\bm{k}\cdot\bm{\rho}}\hk]\right> d\bm{v}\\
\nonumber
&= \sum_{\bm{k}} e^{i\bm{k}\cdot\bm{r}}\left[ 
\int \begin{pmatrix} 1\\ v_\parallel\unit{b}\\ v^2\end{pmatrix}
\left<C[\hk]\right> d\bm{v} 
- i\bm{k}\cdot\bm{\hat\Gamma}_{\bm{k}}^{\mathrm{(coll)}}\right]\\ 
&= - \bm{\nabla}\cdot{\hat\Gamma}^{\mathrm{(coll)}}, 
\label{moments}
\end{align}
where $\bm{\hat\Gamma}^{(0)}$ denotes the fluxes arising from the terms in \eqref{GK_eqn} 
other than the collision operator. 
The moments of the gyroaveraged collision operator have been expressed as the moments 
of the same operator formally taken at $\bm{k}\cdot\bm{\rho}=0$ 
(i.e., dropping all FLR contributions: 
$\left<C[\hk]\right>$ instead of $\left<e^{i\bm{k}\cdot\bm{\rho}}C[e^{-i\bm{k}\cdot\bm{\rho}}\hk]\right>$) 
plus the divergence of the collisional flux arising from the finite-Larmor-radius 
part of $\Cgk[h]$. This representation was achieved by expanding all 
gyrophase factors in \eqref{moments} into infinite Taylor series 
and noticing that they take the form 
$e^{\pm i\bm{k}\cdot\bm{\rho}} = 1 \pm i\bm{k}\cdot\bm{\rho} - (1/2)(\bm{k}\cdot\bm{\rho})^2 + \dots 
= 1 - i\bm{k}\cdot\left[\dots\right]$, where the square brackets contain the rest of the series. 
Without the FLR terms, the particle, momentum and energy moments 
of the gyrokinetic collision operator vanish,
\begin{equation}
\int \begin{pmatrix} 1\\ v_\parallel\\ v^2\end{pmatrix}
\left<C[\hk]\right> d\bm{v} = 0, 
\label{cons_constraint}
\end{equation}
and \eqref{moments} represents the local conservation law for these 
quantities, with the fluxes containing both collisionless and collisional 
contributions: $\bm{\hat\Gamma} = \bm{\hat\Gamma}^{(0)}+\bm{\hat\Gamma}^{\mathrm{(coll)}}$.\footnote{We thank G.~Hammett for suggesting this interpretation of local conservation laws for the gyrokinetic collision operator.} 
Thus, a conservative numerical implementation of the gyrokinetic collision operator 
can be achieved if \eqref{cons_constraint} is hard-wired into the numerical sheme.\footnote{Note that in order to achieve a conservative numerical implementation of the operator \eqref{gyroav}, 
it turns out to be convenient to write the integral field-particle terms in a slightly modified, 
explicitly conservative form using the identities given in \eqref{dnu_def} and \eqref{nuE_def}. 
See Paper II\cite{numerics} for further details.} 
How to do this is explained in Paper II,\cite{numerics} where we also demonstrate the correct performance of our model operator on a number of test problems. 
 
\section{Interspecies Collisions}
\label{sec_ei}

Let us now turn to the collisions between different species and focus on a plasma containing 
only electrons and one species of ions with a mass ratio $m_e/m_i\ll1$. The smallness of the 
mass ratio allows for a significant simplification of the interspecies collision terms. 
Since ion-electron collisions are subdominant to the ion-ion ones,\cite{helander2002ctm} 
$\nu_{ie}/\nu_{ii}\sim(m_e/m_i)^{1/2}$, to lowest order in the mass ratio we can neglect 
the ion-electron collisions and the ion collisions can be modelled using the like-particle
operator proposed above [\eqref{gyroav}].   
The situation is different for the electron-ion collisions, which are same order in 
mass ratio as the electron-electron collisions,\cite{helander2002ctm} $\nu_{ei}\sim\nu_{ee}$. 
Thus, the full electron collision operator has two parts: 
\begin{equation}
C[\delta f_e]=C_{ee}[\delta f_e] + C_{ei}[\delta f].
\end{equation} 
The electron-electron operator $C_{ee}[\delta f_e]$ can be modelled by the like-particle operator proposed above [\eqref{model}], the electron-ion collision operator can be expanded in the mass ratio and to two leading orders reads\cite{helander2002ctm} 
\begin{eqnarray}
C_{ei}[\delta f] &=& \nu_D^{ei} \left(L[\delta f_e] + 
\frac{2 \bm{v}\cdot\bm{u}_i }{{\vth^2}_{e}} F_{0e} \right),\\ 
\nu_D^{ei}(v) &=& \nu_{ei}\left(\frac{{\vth}_e}{v}\right)^3,
\label{Cei}
\end{eqnarray}
where 
$\nu_{ei} = \sqrt{2}\pi n_{0i} Z^2 e^4 \ln \Lambda\,T_{0e}^{-3/2} m_e^{-1/2}$ 
is the dimensional electron-ion collision frequency, $Z=q_i/e$, $e$ is the 
electron charge, 
$L$ is the Lorentz operator given by \eqref{def_L}, and 
\begin{equation}
\bm{u}_i = \frac{1}{n_{0i}} \int \bm{v} \delta f_i\,d\bm{v}
\end{equation}
is the ion flow velocity. Thus, the electron-ion collisions are correctly modelled to lowest order in the mass ratio by electron pitch-angle scattering off static ions, with electron drag against the bulk ion flow as a first order correction to this. 

Note that the drag term is necessary to correctly capture electron-ion friction and hence resistivity; failure to include it leads to incorrect results, with mean electron momentum relaxed towards zero rather than towards equality with the mean ion momentum. The need to include this first-order effect stems from the fact that whilst the effect this has on the mass flow is small (mass is predominantly carried by the ions) there is a large effect on the current. Including the small correction to the mean ion motion due to friction on electrons results in a small correction to both the current and the mass flow. However, these first-order effects such as the small slow collisional change in the mean ion momentum, are formally of the same order as the drag terms in the electron-ion collision operator. We will, therefore, keep the first-order correction to the ion collision operator.

Taking the lowest-order contribution to the linearized ion-electron operator\cite{helander2002ctm}
\begin{align}
C_{ie}[\delta f] = \frac{\bm{R}_{ie}}{m_i n_{0i}} \cdot \pd{F_{0i}}{\bm{v}},
\end{align}
where $\bm{R}_{ie}$ is the ion-electron friction force. 
Since $\bm{R}_{ie} = -\bm{R}_{ei} = -\int m_e\bm{v} C_{ei}[\delta f_e] d\bm{v}$, 
the ion-electron collision operator is expressed in terms of the perturbed electron distribution
function: using lowest-order term in \eqref{Cei} (the pitch-angle scattering), we find, to lowest 
order in the mass-ratio expansion,
\begin{align}
C_{ie}[\delta f_e] =
-\frac{F_{0i}}{n_{0i} T_{0i}}\, \bm{v} \cdot \int \bm{v}' m_e \nu_D^{ei}(v') \delta f_e(\bm{v}')
d\bm{v}'.
\label{Cie}
\end{align}
It is now easy to see that formally the ion-electron collisions must be kept in order 
for the $H$-theorem to be satisfied. Indeed, for the interspecies collisions, the 
$H$-theorem is written as follows 
\begin{align}
\iint \frac{T_{0i}\delta f_i}{F_{0i} } C_{ie} [\delta f] d\bm{v} d\bm{r} 
+\iint \frac{T_{0e}\delta f_e}{F_{0e} } C_{ei} [\delta f] d\bm{v} d\bm{r} 
\le 0
\end{align}
(i.e., the interspecies terms in \eqref{balance} are nonpositive). 
We can see immediately that the pitch-angle-scattering part of the electron-ion operator 
automaticlaly satisfies this, while  
the contribution from the drag term in \eqref{Cei} is exactly cancelled by 
the contribution from the ion-electron operator given by \eqref{Cie}.

Let us now work out the gyrokinetic interspecies collision operators. 
Performing the conversion of the electron-ion operator [\eqref{Cei}] to the gyroaveraged form in a way entirely analogous to what was done in \secref{sec_gk} and \apref{ap_gyroavg}, we get 
\begin{widetext}
\begin{eqnarray}
\nonumber
\Cgk^{ei}[h_{e\bm{k}}] &=& \nu_D^{ei} \left[ 
\frac{1}{2}\,\pd{}{\xi}(1-\xi^2)\pd{h_{e\bm{k}}}{\xi}
- \frac{1}{4}(1 + \xi^2) \frac{v^2}{{\vth^2}_e}\, k_\perp^2 \rho_e^2 h_{e\bm{k}}
+ \frac{2 v_\parallel J_0(a_e) u_{\parallel i\bm{k}} }{{\vth^2}_{e}} F_{0e} 
\right.\\
&&\qquad\qquad\qquad\qquad\qquad\qquad\qquad\left.
- \frac{Zm_e}{m_i}\frac{v_\perp^2}{{\vth^2}_e} \frac{J_1(a_e)}{a_e} F_{0e} 
k_\perp^2 \rho_i^2\,\frac{1}{n_{0i}}\int \frac{2{v'_\perp}^2}{{\vth^2}_i} 
\frac{J_1(a_i')}{a_i'} h_{i\bm{k}}(\bm{v}') d\bm{v}'\right],
\label{Cgk_ei}
\end{eqnarray}
\end{widetext}
where 
\begin{equation}
u_{\parallel i\bm{k}} = \frac{1}{n_{0i}}\int v_\parallel J_0(a_i) h_{i\bm{k}}\,d\bm{v}
\label{upari}
\end{equation} 
and $a_s=k_\perp v_\perp/\Omega_s$ for species $s$ and the rest of the notation 
as in previous sections, except with species indices reintroduced.

Let us estimate the size of the four terms in \eqref{Cgk_ei} at the ion (long) 
and electron (short) scales. The first term (pitch-angle scattering) is always 
important. At the ion scales, $k_\perp\rho_i\sim1$, the third term (parallel ion drag) is 
equally important, while the second term (electron gyrodiffusion) and 
the fourth term are subdominant by a factor of $m_e/m_i$. 
At the electron scales, $k_\perp\rho_e\sim1$, the pitch-angle scattering 
and the electron gyrodiffusion (the first two terms) are both important. 
Since at these scales $k_\perp\rho_i\sim(m_i/m_e)^{1/2}\gg1$, 
the third and fourth terms are subdominant by a factor (resulting from the 
Bessel functions under the velocity integrals) of $1/\sqrt{k_\perp\rho_i}\sim(m_e/m_i)^{1/4}$. 
In fact, they are smaller than this estimate because at these short wavelengths, 
the ion distribution function has small-scale structure in velocity space 
with characteristics scales $\delta v_\perp/{\vth}_i\sim1/k_\perp\rho_i$, 
with leads to the reduction of the velocity integrals by another factor of  
$1/\sqrt{k_\perp\rho_i}$. Thus, at the electron scales, the third and fourth terms 
in \eqref{Cgk_ei} are subdominant by a factor of $(m_e/m_i)^{1/2}$. 

These considerations mean that the fourth term in \eqref{Cgk_ei} is always negligible 
and can safely be dropped. The full model gyrokinetic electron collision operator is, 
therefore, 
\begin{widetext}
\begin{equation}
\label{model_e}
\Cgk[h_{e\bm{k}}] = \Cgk^{ee}[h_{e\bm{k}}] + \nu_D^{ei} \left[ 
\frac{1}{2}\,\pd{}{\xi}(1-\xi^2)\pd{h_{e\bm{k}}}{\xi}
- \frac{1}{4}(1 + \xi^2) \frac{v^2}{{\vth^2}_e}\, k_\perp^2 \rho_e^2 h_{e\bm{k}}
+ \frac{2 v_\parallel J_0(a_e) u_{\parallel i\bm{k}} }{{\vth^2}_{e}} F_{0e} \right],
\end{equation}
\end{widetext}
where the electron-electron model operator $\Cgk^{ee}[h_{e\bm{k}}]$ is given by \eqref{gyroav} 
and $u_{\parallel i\bm{k}}$ by \eqref{upari}. 
Note that since $e n_{0e}(u_{\parallel i}-u_{\parallel e}) = j_\parallel$ is the parallel current, 
the parallel Amp\`ere's law can be used to express $u_{\parallel i}$ in \eqref{model_e} 
in a form that does not contain an explicit dependence on the ion distribution function: 
\begin{equation}
u_{\parallel i\bm{k}} = \frac{1}{n_{0e}}\int v_\parallel J_0(a_e)h_{e\bm{k}}\, d\bm{v} 
+ \frac{c}{4\pi e n_{0e}}\,k_\perp^2 A_{\parallel\bm{k}}.
\end{equation}
This turns out to be useful in the numerical implementation of the electron operator, 
detailed in Paper II.\cite{numerics}

Performing a completely similar calculation for the ion-electron operator [\eqref{Cie}], 
we obtain the full model gyroaveraged ion collision operator: 
\begin{widetext}
\begin{equation}
\label{model_i}
\Cgk[h_{i\bm{k}}] = \Cgk^{ii}[h_{i\bm{k}}] - \frac{F_{0i}}{n_{0i} T_{0i}}\,v_\parallel J_0(a_i) 
\int v_\parallel' J_0(a_e') m_e \nu_D^{ei}(v') h_{e\bm{k}}(\bm{v}') d\bm{v}',
\end{equation}
\end{widetext}
where the ion-ion model operator $\Cgk^{ee}[h_{e\bm{k}}]$ is given by \eqref{gyroav}.

\section{Summary}
\label{sec_summary}

Thus in \secref{sec_intro} we have argued the necessity of dissipation in turbulence simulations, justified the direct modelling of collisions in order to provide such dissipation and postulated a set of constraints for a physically reasonable model collision operator. Previously used model operators mostly do not contain energy diffusion and were thus deemed unsatisfactory for these purposes.
Of the existing model operators that contain energy diffusion, two are detailed in Ref.\ \onlinecite{catto1977col} and Ref.\ \onlinecite{hirshman1976afp}. The former however does not satisfy the $H$-theorem [\eqref{Htheorem}] and the latter incorrectly captures the smallest scales. These problems are demonstrated and discussed in detail in \apref{ap_prev}. 

In \secref{sec_newop} we presented a new operator [\eqref{model}] that successfully introduces energy diffusion whilst maintaining the $H$-Theorem and conservation laws, thus satisfying the conditions set forth in the introduction. This operator is then transformed into gyrokinetic form in \secref{sec_gk} correctly accounting for the gyrodiffusive terms and FLR effects [\eqref{gyroav}]. In order to provide a complete recipe for modelling the collisional effects in simulations, the same gyroaveraging procedure is applied in \secref{sec_ei} to electron--ion and ion--electron collisions, somewhat simplified by the mass-ratio expansion [\eqref{model_e} and \eqref{model_i}]. This leaves us with a complete picture of collisions in gyrokinetic simulations, capturing gyrodiffusion, resitivity and small-scale energy diffusion.

When we discussed the gyroavergaing procedure in \secref{sec_gk} we presented the specific case of the application to Eulerian flux-tube $\delta f$ gyrokinetic simulations.\cite{jenko:1904,candy2003egm} However, the form presented in \eqref{model} is suitable for inclusion in most $\delta f$ kinetic systems and even amenable to use in Lagrangian codes by applying the methods of Refs.~\onlinecite{xu1991nsi} or \onlinecite{dimits1994cop} to the gyroaveraged operator given by \eqref{gyroav}. Indeed, by suitable discretization of the gyroaveraging procedure\cite{candy2003egm} it would also be usable in a global Eulerian code.

We conclude by noting that the final arbiter of the practicality and effectiveness of this collision model is the numerical implementation and testing performed in Paper II,\cite{numerics} where our operator is integrated into the \verb#GS2# code. The battery of tests shows that our operator not only reproduces the correct physics in the weakly collisional regime but even allows a gyrokinetic code to capture correctly the collisional (reduced-MHD) limit.

\begin{acknowledgments} 
We thank G.\ Hammett, J.\ Hastie, D.\ Ernst, P.\ Ricci, C.\ Roach, B.\ Rogers and T.\ Tatsuno for useful discussions. G.\ Hammett has also made several suggestions on the manuscript that have helped improve both the substance and the style of the presentation. I.G.A.\ was supported by a CASE EPSRC studentship in association with UKAEA Fusion (Culham). M.B.\ was supported by the US DOE Center for Multiscale Plasma Dynamics. A.A.S. was supported by an STFC (UK) Advanced Fellowship and STFC Grant ST/F002505/1. M.B.\ and W.D.\ would also like to thank the Leverhulme Trust (UK) International Network for Magnetized Plasma Turbulence for travel support.
\end{acknowledgments}

\appendix
\section{Proof of The $H$-Theorem for \eqref{model}}
\label{ap_Hth}
In the case of the expansion $f=F_0 + \delta f$ about a Maxwellian the entropy generation by like particle collisions takes the form
\begin{eqnarray}
\nonumber
\frac{d S}{dt} &=& - \frac{d}{dt} \iint f \ln f \,d\bm{v} \,d\bm{r}\\
&=& - \iint \hat{f} {C}[\hat{f} F_0] \,d\bm{v}\,d\bm{r}\ge0,
\label{entropy}
\end{eqnarray}
where we use the compact notation $\hat{f} = \delta f / F_0$.
The statement of the $H$-theorem is that the right-hand side of \eqref{entropy} is nonnegative 
and that it is exactly zero when $\delta f$ is a perturbed Maxwellian. 

We represent $\hat{f}$ as a Cartesian tensor expansion (or equivalently spherical harmonic expansion) in velocity space:
\begin{eqnarray}
	\hat{f}(\bm{r},\bm{v}) = \hat{f}_{0}(\bm{r},v) + \bm{v} \cdot \bm{\hat{f}}_{1}(\bm{r},v) + {R}[\hat{f}](\bm{r},\bm{v}),
\label{expansion}
\end{eqnarray}
where $R[\hat{f}]$ comprises the higher order terms. 
It is then possible to recast the statement of the $H$-Theorem in terms of this expansion using linearity of the model collision operator $C$ [\eqref{model}], orthogonality of the expansion and the fact that spherical harmonics are eigenfunctions of the Lorentz operator ${L}$. 
By construction, $R[\hat{f}]$ satisfies 
$\int R[\hat{f}]F_0 {d}\bm{v} = 0$ and 
$\int \bm{v} R[\hat{f}] F_0{d}\bm{v} = \bm{0}$, from which it follows that
$R[\hat{f}]$ does not contribute to the field-particle parts of the model operator:
$\bm{U}[ R [ \hat{f} ]F_0 ] = \bm{0}$ and $Q[ R [ \hat{f} ] F_0] = 0$. 
Substituting \eqref{expansion} into the right-hand side of \eqref{entropy}, where the operator $C$ is given by \eqref{model}, and integrating by parts those terms involving derivatives of $R[\hat{f}]$, we find that they all give nonnegative contributions, so we have 
\begin{eqnarray}
 - \int\hat{f} {C}[\hat{f}F_0] d\bm{v} \ge \sigma_0 + \sigma_1,
\end{eqnarray}
where 
\begin{align}
\label{sigma0_def}
\sigma_0 &= -\int \hat{f}_0 C[\hat{f}_0 F_0]d\bm{v},\\
\sigma_1 &= -\int \bm{v} \cdot \bm{\hat{f}}_1 {C}[\bm{v}\cdot\bm{\hat{f}}_1 F_0] d\bm{v}.
\label{sigma1_def}
\end{align}
In order to prove the $H$-theorem, 
it is now sufficient to show that $\sigma_0 \ge 0$ and $\sigma_1 \ge 0$. 

Starting with $\sigma_0$ and using \eqref{model}, we integrate over angles and 
use the differential identity given in \eqref{nuE_def} to express 
the term containing $Q$: 
\begin{equation}
\sigma_0 = -2\pi \int \hat{f}_0\pd{}{v}\left[ v^4 \nu_{\parallel}F_{0} \pd{}{v}
\left( \hat{f_0} - \frac{v^2}{\vth^2}\,Q[\hat{f}_0 F_0] \right) \right] dv.
\end{equation}
Using the aforementioned identity again in the expression for $Q$ [\eqref{def_Q}] and 
integrating by parts where opportune, we get
\begin{equation}
\begin{split}
\sigma_0 &= 4\pi\left[\frac{1}{2}\int\left(\pd{\hat{f}_0}{v}\right)^2 v^4\nu_\parallel F_0dv 
\right.\\
&- \left.\left.\left(\int \pd{\hat{f}_0}{v} v^5\nu_\parallel F_0 dv\right)^2
\right/ \int v^6\nu_E F_0 dv
\right].
\end{split}
\label{sigma0}
\end{equation}
It is easy to see from the Cauchy-Schwarz inequality that
\begin{equation}
\label{sigma0_CS}
\left(\int{ \pd{\hat{f}_{0}}{v} v^5 \nu_\parallel F_{0} dv } \right)^2 \le 
\int \left(\pd{\hat{f}_{0}}{v} \right)^2 v^4 \nu_\parallel F_{0} dv 
\int v^6 \nu_\parallel F_{0} dv.
\end{equation}
Using this in the second term of \eqref{sigma0}, we infer
\begin{equation}
\begin{split}
\sigma_0 &\ge 4\pi \int \left(\pd{\hat{f}_{0}}{v} \right)^2 v^4 \nu_\parallel F_{0} dv\\
&\times\left(\frac{1}{2}-\left.\int v^6\nu_\parallel F_0 dv\right/ \int v^6\nu_E F_0 dv\right)
= 0, 
\end{split}
\label{sigma0_ineq}
\end{equation}
where to prove that the right-hand side vanishes, we  
again used the differential identity given in \eqref{nuE_def}
and integrated by parts. 
Thus, we have proved that $\sigma_0 \ge 0$. 

Turning now to $\sigma_1$ [\eqref{sigma1_def}], using \eqref{model}, and integrating 
by parts where opportune, we get
\begin{equation}
\begin{split}
\sigma_1 &= \int ( \bm{v}\cdot\bm{\hat{f}}_1 )^2 \nu_D F_{0} d\bm{v} 
- \frac{1}{2} \int \left(\pd{}{v} \bm{v}\cdot\bm{\hat{f}}_1 \right)^2 v^2 \nu_\parallel F_{0} d\bm{v}\\ 
&\quad- 3 \vth^2 \left.\left|\int \bm{x} \bm{x}\cdot\bm{\hat{f}}_1 \nu_s F_{0} d\bm{v} \right|^2 \right/ \int x^2 \nu_s F_{0} d\bm{v}, 
\end{split}
\end{equation}
where we have used the standard notation that $\bm{x} = \bm{v} / \vth$ and $x = v/\vth$. 
Integrating over angles and using the simple identity $\bm{a} \cdot \int \hat{\bm{v}}\hat{\bm{v}} d\Omega = (4\pi/3)\bm{a}$, where $\hat{\bm{v}}=\bm{v}/v$ and $\bm{a}$ is an arbitrary vector, we have
\begin{equation}
\begin{split}
\sigma_1 &= \frac{4\pi\vth^5}{3} \Biggl( \int\! |\bm{\hat{f}}_1|^2 x^4 \nu_D F_{0} dx
+ \frac{1}{2} \int \left| \pd{}{x}\,x\bm{\hat{f}}_1\right|^2\!\! x^4 \nu_\parallel F_{0} dx \Biggr. \\  
&\quad - \left.\left.\left| \int \bm{\hat{f}}_1 x^4 \nu_s F_{0} dx\right|^2\right/ \int x^4 \nu_s F_{0} dx  
\right).
\end{split}
\label{sigma1}
\end{equation}
Once again applying the Cauchy-Schwarz inequality, we find that
\begin{equation}
\left| \int \bm{\hat{f}}_1 x^4\nu_s F_{0} dx \right|^2 \le 
\int \left| \bm{\hat{f}}_1 \right |^2 x^4 \nu_s F_{0} dx
\int x^4 \nu_s F_{0} dx.
\end{equation}
Using this in the last term in \eqref{sigma1}, we get 
\begin{equation}
\begin{split}
\sigma_1   \ge  \frac{4\pi}{3}\vth^5
&\left( \int\left|\bm{\hat{f}}_1\right|^2 x^4 \Delta\nu F_{0} {d}x \right.\\
&+ \left.\frac{1}{2} \int \left|\pd{}{x}\,x\bm{\hat{f}}_1\right|^2 
x^4\nu_\parallel F_{0} dx\right),
\end{split}
\end{equation}
where $\Delta\nu$ is defined in \eqref{dnu_def}. 
Upon using the differential identity 
given in \eqref{dnu_def} to express $\Delta\nu$ 
in the first term of the above expression and integrating the resulting expression by parts, 
we finally obtain
\begin{eqnarray}
\sigma_1 \ge \frac{4\pi}{3}\vth^5
\int \left|\pd{\bm{\hat{f}}_1}{x} \right|^2 x^6\nu_\parallel F_{0} {d}x\ge0.
\end{eqnarray}

We now consider when these inequalities becomes equalities, i.e., 
when the right-hand side of \eqref{entropy} is zero. 
Firstly, this requires $\partial R[\hat{f}]/\partial \xi = 0$ and thus $R[\hat{f}] = 0$, 
so $\hat{f}$ contains no 2nd or higher-order spherical harmonics. 
Secondly, $\sigma_0 = 0$ if either $\hat{f}_0$ is independent of $v$ or we have equality in the invocation of the Cauchy-Schwarz inequality [\eqref{sigma0_CS}], which occurs if $\hat{f}_0 \propto v^2$. Similarly $\sigma_1 = 0$ iff $\bm{\hat{f}}_1$ is independent of $v$. 
Thus, the right-hand side of \eqref{entropy} 
vanishes iff $\hat{f} \propto 1, \bm{v}, v^2$, i.e., $\delta f= \hat{f} F_0$ 
is a perturbed Maxwellian.   

This completes the proof of the $H$-theorem for our model operator. 

\section{ Gyroaveraging }
\label{ap_gyroavg}

To transform the derivatives in \eqref{model} from the original phase-space coordinates 
$(\bm{r},v,\xi,\vartheta)$ to the new coordinates $(\bm{R},v,\xi,\vartheta)$, 
we require the following formulae:
\begin{align}
\left(\pd{}{{v}}\right)_{\bm{r}} &= \left(\pd{}{{v}}\right)_{\bm{R}} 
- \frac{1}{v}\,\bm{\rho}\cdot\left(\pd{}{\bm{R}}\right)_{\bm{v}},\\
\left(\pd{}{\xi}\right)_{\bm{r}} &= \left(\pd{}{\xi}\right)_{\bm{R}} 
+ \frac{\xi}{1-\xi^2}\,\bm{\rho} \cdot \left(\pd{}{\bm{R}}\right)_{\bm{v}}, \\
\left(\pd{}{\vartheta}\right)_{\bm{r}} &= \left(\pd{}{\vartheta}\right)_{\bm{R}} 
+ \frac{\bm{v}_\perp}{\Omega}\cdot \left(\pd{}{\bm{R}}\right)_{\bm{v}},
\end{align}
where $\bm{\rho}=\unit{b}\times\bm{v}_\perp/\Omega$. 
In Fourier-transformed perpendicular guiding center variables, we can replace 
in the above formulae $(\partial/\partial\bm{R})_{\bm{v}}\to i\bm{k}$, where 
$\bm{k}\equiv\bm{k}_\perp$. It is convenient to align (without loss of generality) 
the $\vartheta=0$ axis with $\bm{k}$, so we have 
$\bm{v}_\perp\cdot\bm{k} = k_\perp v\sqrt{1-\xi^2}\cos\vartheta$ 
and $\bm{\rho}\cdot\bm{k} = -k_\perp v\sqrt{1-\xi^2}\sin\vartheta$. 
Using the above formulae, we gyroaverage the Lorentz operator in \eqref{model}:
\begin{equation}
\label{avg_L}
\left<L[\hk]\right> = {1\over2}\pd{}{\xi}(1-\xi^2)\pd{\hk}{\xi}
- \frac{v^2(1+\xi^2)}{4\Omega^2}k_\perp^2\hk,
\end{equation}
where we have used $\gyroR{\bm{\rho}\bm{\rho}} = \gyroR{\bm{v}_\perp\bm{v}_\perp}=(1/2)\bm{I}$. 
Note that both the terms containing $\xi$ and $\vartheta$ derivatives in the 
original operator \eqref{def_L} produce non-zero gyrodiffusive contributions 
[the second term in \eqref{avg_L}]. Another such gyrodiffusive term, equal to 
$-\nu_\parallel \left[v^2(1-\xi^2)/4\Omega\right]k_\perp^2\hk$, arises from the 
energy-diffusion part of the test-particle operator in \eqref{model}. 
Collecting these terms together and defining the thermal Larmor 
radius $\rho=\vth/\Omega$, we arrive at the gyrodiffusion term in 
\eqref{gyroav}. 

It remains to gyroaverage the field-particle terms. 
For the energy-conservation term [\eqref{def_Q}] we have
\begin{equation}
\left<e^{i\bm{k}\cdot\bm{\rho}}Q[e^{-i\bm{k}\cdot\bm{\rho}}\hk]\right> = 
\left<e^{i\bm{k}\cdot\bm{\rho}}\right>Q[e^{-i\bm{k}\cdot\bm{\rho}}\hk], 
\label{avgQ}
\end{equation}
where 
\begin{equation}
\begin{split}
&Q[e^{-i\bm{k}\cdot\bm{\rho}}\hk] = \\
&\qquad\int v^2 \nu_E \left<e^{-i\bm{k}\cdot\bm{\rho}}\right>\hk d\bm{v}
\left/ \int v^2 \left(v/\vth\right)^2 \nu_E F_{0} d\bm{v}. \right.
\end{split}
\label{Qhk_ap}
\end{equation}
Note that the $\vartheta$ integration in $Q$ only affected $e^{-i\bm{k}\cdot\bm{\rho}}$, 
hence the above expression. 
Using the standard Bessel function identity\cite{watson1966ttb}
$\int\limits_0^{2\pi} e^{i a \sin \vartheta } \,d\vartheta = 2\pi J_0(a)$,
we find $\left<{e^{-i\bm{k}\cdot\bm{\rho}}}\right> = J_0(a)$, where $a=k_\perp v_\perp/\Omega$.
Substituting this into \eqsref{avgQ}{Qhk_ap}, we arrive at the energy-conservation term 
in \eqref{gyroav}, where the expression in the right-hand side of \eqref{Qhk_ap} is denoted 
$Q[\hk]$ [\eqref{Qhk}].  

The momentum-conserving terms are handled in an analogous way: details can be found in Appendix\ B of Ref.\ \onlinecite{Tome}, where a simpler model operator was gyroaveraged.

\section{ Comparison with previous model opreators }

\label{ap_prev}
In order to compare and contrast with previously suggested operators that do include energy diffusion, we first rewrite in our notation the operator derived by Catto and Tsang 
[Eqs.~(14) and (16) in Ref.\ \onlinecite{catto1977col} ],
\begin{widetext}
\begin{equation}
{C}_\mathrm{CT}\left[\delta f\right] = \nu_D {L}[ \delta f ] 
+ \frac{1}{v^2} \pd{}{v}\left(\frac{1}{2}\, v^4 \nu_\parallel F_{0} \pd{}{v}\frac{\delta f}{F_0} \right)
+ \frac{2 F_0}{n_{0} \vth^2 }\,\bm{v} \cdot \int \bm{v} \nu_s\delta f d\bm{v}
+ \frac{2 F_0}{3n_{0}} \left( \frac{v^2}{\vth^2}-\frac{3}{2}\right)
\int \frac{v^2}{\vth^2}\, \nu_E\delta f d\bm{v}. 
\label{CT_op}
\end{equation}
\end{widetext}
This operator, whilst it conserves particle number, momentum and energy, neither obeys the $H$-Theorem nor vanishes on a perturbed Maxwellian. 

The latter point can be demonstrated most easily by letting $\delta f = x^2F_0$, 
where $x=v/\vth$. This $\delta f$ is proportional to a perturbed Maxwellian with non-zero $\delta n$ and $\delta T$. We can then evaluate the test-particle and field-particle parts of the operator to find
\begin{equation}
\frac{1}{v^2} \pd{}{v}\left( \frac{1}{2} v^4 \nu_\parallel F_{0} \pd{}{v}\frac{\delta f}{F_0} \right) 
= \frac{1}{x^2} \pd{}{x} \left( x^5 \nu_\parallel F_{0} \right) = - x^2 \nu_E F_{0}
\end{equation}
and
\begin{equation}
\int x^2 \nu_E \delta f d\bm{v} 
= \frac{4 n_{0}}{\sqrt{\pi}} \int\limits_0^{\infty} x^6 \nu_E e^{-x^2} {d}x 
= \sqrt\frac{2}{\pi} \,n_0\nu.
\end{equation}
Substituting into \eqref{CT_op}, we get
\begin{equation}
{C}_\mathrm{CT}[\delta f] = -x^2\nu_E F_{0} + \sqrt\frac{2}{\pi} 
\left(x^2-\frac{3}{2}\right) \nu F_{0},
\end{equation}
which is non-zero despite $\delta f$ being a perturbed Maxwellian. 

In order to show that the $H$-theorem can be violated by the operator \eqref{CT_op}, 
let us consider a perturbed distribution function of the form $\delta f = x^3 F_0$. Then
\begin{equation}
C_{\mathrm{CT}}[\delta f] = \left[ \frac{3}{2} \left(\nu_\parallel -\nu_E\right) + \left(x^2 - \frac{3}{2}\right) \nu \right] F_0,
\end{equation}
so the entropy generation is,
\begin{equation}
\frac{dS}{dt} = - \iint \frac{\delta f}{F_0}\,{C}_\mathrm{CT}\left[\delta f\right] d\bm{v}d\bm{r}
=-\frac{3}{64}(32-21\sqrt{2})\,\nu V < 0, 
\end{equation}
where $V$ is the volume of the system. The above expression is negative, which breaks the 
$H$-theorem and produces unphysical plasma cooling for the particular form of the perturbed 
distribution function that we have examined. 

The second case we examine here is the sequence of operators derived by Hirshman and Sigmar.\cite{hirshman1976afp} The general operator proposed by these authors is given in their Eq.~(25). In our notation, we rewrite here the $N=0$, $l=0,1$ restriction of the like-particle form of their operator with $\Delta \nu$ set to $0$ for simplicity (this does not affect the discussion that follows): 
\begin{widetext}
\begin{equation}
\begin{split}
	C_\mathrm{HS}[\delta f] 
	&=
	\nu_D {L}[\delta f] + \frac{1}{v^2} \pd{}{v}
	\left[
			\frac{1}{2} v^4 \nu_{\parallel} F_{0} \pd{}{v} 
				\left(
					\frac{1}{4\pi}\int\limits_{-1}^{1} {d} \xi \int\limits_{0}^{2\pi} {d} \vartheta \frac{\delta f}{F_0} - \frac{v^2}{\vth^2}\,Q[\delta f]
				\right)
	\right]
	+ \nu_D\, \frac{2\bm{v} \cdot \bm{U}}{\vth^2} F_0, 
\end{split}
\end{equation}
where $\bm{U}$ and $Q$ are defined by \eqsref{def_U}{def_Q}. 
The primary concern here comes from the angle averaging operation in the energy-diffusion part of the operator. Firstly, the energy diffusion only acts on the spherically symmteric (in velocity space) part of the perturbed distribution function. However, there is no reason why there cannot arise perturbations that have very large energy derivatives but angle-average to zero (for example, $\delta f\propto\xi$). Clearly, such perturbations will not damped correctly. Secondly, upon conversion to gyrokinetic coordinates and gyroaveraging (see \secref{sec_gk} and \apref{ap_gyroavg}), the operator becomes
\begin{equation}
\begin{split}
C_\mathrm{HS,GK}[\hk] &=
\nu_D(v) \left[ \frac{1}{2} \frac{\partial}{\partial \xi} \left(1-\xi^2\right) \pd{\hk}{\xi} 
- \frac{1}{4}(1+\xi^2)\frac{v^2}{\vth^2}\,k_{\perp}^2\rho^2 \hk \right] 
+ \frac{J_{0}(a)}{v^2}\pd{}{v}\left[ \frac{1}{2}\, v^4 \nu_\parallel F_{0}
\pd{}{v}\int\limits_{{-1}}^{1} {d\xi} \frac{J_0(a) \hk}{2\pi F_{0}} \right]\\
&\qquad + 2\nu_D\frac{v_\perp J_1(a) U_\perp \left[ \hk \right] +  v_\parallel J_0(a) U_\parallel \left[ \hk \right] }{ \vth^2} F_{0} + \nu_E\, \frac{v^2}{\vth^2}\,J_0(a)Q[\hk] F_{0},
\end{split}
\end{equation}
\end{widetext}
where the conservation functionals $U_\perp$, $U_\parallel$ and $Q$ are the same as defined in \eqsdash{Uperp}{Qhk}. The immediatly obvious problem is that the angle averaging has introduced two new Bessel functions into the energy diffusion term. The energy diffusion is therefore supressed by one power of $k_\perp \rho$ in the limit $k_\perp \rho\gg1$, while it is precisely in this limit that we expect the small-scale structure in the velocity space to be particularly important.\cite{Tome,schekcrete} This means that the energy cutoff in phase space is artificially pushed to smaller scales and one might encounter all the problems associated with insufficient energy diffusion.\cite{numerics}

While, for the reasons outlined above, we expect the Hirshman--Sigmar operator not to be a suitable 
model for collisions, we would like to note that for many purposes the Hirshman-Sigmar operators are superior to the model operator we presented in \secref{sec_newop}. Taken as a sequence, they provide a rigorous way of obtaining classical and neoclassical transport coefficients to any desired degree of accuracy, and it is relatively easy to solve the Spitzer problem for them, while the Spitzer functions for our operator are hard to find analytically.

\end{document}